\documentclass{article}
\usepackage{mathtools}
\usepackage{graphicx}
\usepackage{amsmath}
\usepackage{amssymb}
\usepackage{amsthm}
\usepackage{iftex}
\usepackage{lmodern}
\usepackage{textcomp}
\usepackage{multirow}
\usepackage[symbol]{footmisc}

\theoremstyle{thmstyleone}
\newtheorem{theorem}{Theorem}[section]
\newtheorem{proposition}{Proposition}[section]
\newtheorem{example}{Example}[section]%
\newtheorem{remark}{Remark}[section]%
\newtheorem{lemma}{Lemma}[section]%

\begin{document}
\begin{center}
	\textbf{\huge MDS and MHDR cyclic codes over finite chain rings}
\end{center}
\begin{center}
	Monika Dalal$^{1}$, Sucheta Dutt$^{1*}$ and Ranjeet Sehmi$^{1}$\\
$^{1}$Department of Applied Sciences,Punjab Engineering College\\
 (Deemed to be University), Chandigarh, India, 160012\\
$^{*}$Corresponding author(s). E-mail(s): $sucheta@pec.edu.in$ ;\\
Contributing authors : $monika.phdappsc@pec.edu.in$;\\
$rsehmi@pec.edu.in$; 
\end{center}





\section*{Abstract}\label{sec0}

       In this work, a unique set of generators for a cyclic code over a finite chain ring has been established. The minimal spanning set and rank of the code have also been determined. Further, sufficient as well as necessary conditions for a cyclic code to be an MDS code and for a cyclic code to be an MHDR code have been obtained. Some examples of optimal cyclic codes have also been presented.\\\\
       \textbf{Keywords :} Cyclic codes, Generators, Uniqueness, MDS, MHDR\\
   

\section{Introduction}\label{sec1}
       
          Coding theory aims to provide optimal codes for detecting and correcting maximum number of errors during data transmission through noisy channels. Cyclic codes have been in focus due to their rich algebraic structure which enables easy encoding and decoding of data through the process of channel coding. Cyclic codes over rings have gained a lot of importance after the remarkable breakthrough given by Calderbank et al. in \cite{1}. A vast literature is available on the structure of cyclic codes over fields, integer residue rings, Galois rings and finite chain rings \cite{2,3,4,5,6,7,8,9,10,11,12,13,14,15,16,17,18,19,20,21,22,23,24}
          . Cyclic codes over finite chain rings with length coprime to the characteristic of residue field have been inestigated in \cite{2,18,12}. 
          A. Sharma and T. Sidana have studied cyclic codes of $p^{s}$ length over finite chain rings in \cite{11} thereby extending the results of Kiah et al. on cyclic codes over Galois rings \cite{10}. Dinh et al. have explored the structure and properties of cyclic codes of length $p^{s}$ over finite chain rings with nilpotency index 2 \cite{9}.  However, in most of the studies, there have been some limitation on either the length of code or the nilpotency index of the ring. We do not impose any such restriction in this paper. 
          Salagean made use of the existence of a Grobner basis for an ideal of a polynomial ring to establish a unique set generators for a cyclic code over a finite chain ring with arbitrary parameters \cite{14}. Ashker et al. have also worked in the same direction in the paper \cite{24} by extending the novel approach given by T. Abualrub \cite{8} which pulls back the generators of a cyclic code over $Z_{_{2}}$ to establish the structure of cyclic codes over the ring $Z_{_{2}}+uZ_{_{2}}+\cdots+u^{k-1}Z_{_{k-1}},$ $u^{k}=0.$  They have also extended this approach over the finite chain ring $F_{_{q}}+uF_{_{q}}+\cdots+u^{k-1}F_{_{q-1}},$ $u^{k}=0$ \cite{20}.  Monika et al. have given a constructive approach to establish a generating set for a cyclic code  over a finite chain ring by making use of minimal degree polynomials of certain subsets of the code \cite{16}. We make some advancements to this study by establishing a unique set of generators for a cyclic code over a finite chain ring with arbitrary parameters. It is noted that this unique set of generators retains all the properties of generators obtained in \cite{16}.\\
       
          The manuscript is organised as follows: In section 2, we state some preliminary results. In  section 3, we establish a unique set of generators for a cyclic code over a finite chain ring. In section 4, we establish a minimal spanning set and rank of the cyclic code. We give sufficient as well as necessary conditions for a cyclic code to be an MDS code. We establish sufficient as well as necessary conditions for a cyclic code of length which is not coprime to the characteristic of residue field of the ring, to be an MHDR code. Lastly, we provide a few  examples of MDS and MHDR cyclic codes over some finite chain rings.

\section{Preliminaries}

         Let $R$ be a finite commutative chain ring. Let $\langle \gamma \rangle$  be the unique maximal ideal of $R$ and $\nu$ be the nilpotency index of $\gamma$. Let $F_{q}=R/\langle \gamma \rangle$ be the residue field of $R$, where $q=p^{s}$ for a prime $p$ and a positive integer $s$.\\

        The following is a well known result. For reference, see \cite{11}.

        \begin{proposition}
        	Let $R$ be a finite commutative chain ring. Then
        	\begin{itemize}
        		\item[$(i)$] $charR=p^{a}$, where $1 \leq a \leq \nu$ and $\lvert R \rvert=\lvert F_{q}\rvert^{^{\nu}}=p^{s\nu}$.
        		\item[$(ii)$] There exists an element $\zeta \in R$ with multiplicative order $p^{s}-2$. The set $\top=\{0,1,\zeta,\zeta^{^{2}},\cdots,\zeta^{^{p^{s}-2}}\}$ is called the Teichm$\ddot{u}$ller set of $R$.
        		\item[$(iii)$] Every $r \in R$ can be uniquely expressed as $r=r_{_{0}} + r_{_{1}}\gamma + \cdots + r_{_{\nu-1}}\gamma^{_{\nu-1}}$, where $r_{_{i}} \in \top$ for $0 \leq i \leq \nu-1$. Also, $r$ is a unit in $R$ if and only if $r_{_{0}} \neq 0$.
        	\end{itemize}
        \end{proposition}

        \begin{remark}\label{unique repn of polynomials in R[z]}
        	Let $k(z)=k_{_{0}} + k_{_{1}}z + \cdots + k_{_{t}}z^{t},$ where $k_{_{j}} \in R$ for $j=0,1, \cdots, t$ be a polynomial of degree $t$ in $R[z]$.  Using Proposition $2.1(iii),$ $k(z)$ can be expressed as $$k(z) = a_{_{0}}(z) + \gamma a_{_{1}}(z) + \cdots + \gamma^{\nu-1}a_{_{\nu-1}}(z),$$ where $a_{_{j}}(z) \in \top[z]$ for $j=0,1, \cdots, \nu-1$.
        \end{remark}


          Define a map $\phi : R \rightarrow \top$ by $\phi(r)=r(mod \gamma)=\overline{r}$ for $r \in R$. Clearly $\phi$ is a natural onto homomorphism and therefore $\overline{R}=\top$, where $\overline{R}$ denotes the image of $R$ under $\phi$. This map can be naturally extended from $R[z]$ to $\top[z]$ by $\Sigma_{i=0}^{k}a_{_{i}}z^{i} \mapsto \Sigma_{i=0}^{k}\overline{a_{_{i}}}z^{i}$, where $a_{i} \in R$ for $0 \leq i \leq k$.\\
        

          Let us now recall some basic definitions and known results.\\

          A linear code $C$ with length $n$ over a finite commutative chain ring $R$ is said to be a cyclic code if $(c_{_{n-1}},c_{_{0}},\cdots,c_{_{n-2}}) \in C$ for every $(c_{_{0}},c_{_{1}},\cdots,c_{_{n-1}}) \in C$. It is well established that $C$ can be viewed as an ideal of $R[z]/\langle z^{n}-1 \rangle$. The Hamming weight $w_{_{H}}(c)$ of $c=(c_{_{0}}, c_{_{1}} \cdots c_{_{n-1}}) \in C$ is defined as the number of integers $i$ such that $c_{_{i}}\neq 0$ for $0 \leq i \leq n-1.$ The Hamming distance $d_{_{H}}(C)$ of a code $C$ over $R$ is given by $d_{_{H}}(C)=min\{w_{_{H}}(c) : c ~is ~a ~non$-$trivial ~element ~of ~C \}.$ $C$ is said to be an MDS code with respect to the Hamming metric if $\lvert C\lvert=\lvert R\lvert^{n-d_{H}(C)+1}.$ Rank of $C$ is defined as the total number of elements in the minimal spanning set of $C.$ $C$ is said to be an MHDR code if $d_{_{H}}(C)=n-rank(C)+1.$ The $i^{th}$ torsion code of $C$ is defined as Tor$_{_{i}}(C)=\{\phi(k(z)) \in \overline{R}[z] : \gamma^{i}k(z) \in C\},$ where $0 \leq i \leq \nu-1.$ Then Tor$_{_{i}}(C)$ for all $i,$ $0 \leq i \leq \nu-1$ is a principally generated cyclic code over the residue field of $R.$ The degree of the generator polynomial of Tor$_{_{i}}(C)$ is called the $i^{th}$ torsional degree of $C$. A polynomial in $R[z]$ is said to be monic if its leading coefficient is a unit in $R.$ \\\\

\section{Unique set of generators}

         In this section, a unique set of generators for a cyclic code $C$ of arbitrary length $n$ over $R$ has been established. For this, let us first recall the construction given by Monika et al. to obtain a generating set for a cyclic code $C$ over a finite chain ring $R$ \cite{16}. Let $f_{_{0}}(z),f_{_{1}}(z), \cdots,  f_{_{m}}(z)$ be minimal degree polynomials of certain subsets of $C$ such that $deg\big( f_{_{j}}(z) \big)=t_{_{j}} <n$ and leading coefficient of $f_{_{j}}(z)$ equal to $\gamma^{i_{_{j}}}u_{_{j}},$ where $u_{_{j}}$ is some unit in $R,$ $t_{_{j}} < t_{_{j+1}},$ $i_{_{j}}>i_{_{j+1}}$ and $i_{_{j}}$ is the smallest such power. If $i_{_{0}}=0,$ then $f_{_{0}}(z)$ is monic and we have $m=0.$

        \begin{lemma}[\cite{16}]
       	 Let $C$ be a cyclic code having length $n$ over $R$ and $f_{_{j}}(z)$, $0 \leq j \leq m$ be polynomials as defined above. Then 
       	 \begin{itemize}
       		\item[$(i)$] $C$ is generated by the set $\{ f_{_{j}}(z) ; j=0,1,\cdots,m \}$.
       		\item[$(ii)$]For $0 \leq j \leq m$, $f_{_{j}}(z) = \gamma^{i_{_{j}}} h_{_{j}}(z),$   where $h_{_{j}}(z)$ is a monic polynomial over the finite  commutative chain ring having nilpotency index $\nu - i_{_{j}}$ and maximal ideal $\langle \gamma\rangle.$
       		\item[$(iii)$] $\{ f_{_{j}}(z) ; j=0,1,\cdots,m \}$ forms a Grobner basis for $C.$
       	 \end{itemize}
        \end{lemma}

        The following results are straight forward generalisations of \cite{15} for cyclic codes over the class of Galois rings  to finite chain rings and have been communicated in \cite{17}. These results are required to proceed further.

        \begin{lemma}[\cite{17}]\label{torsion code generators}
        	Consider a cyclic code $C$ of arbitrary length $n$ over $R$ generated by $\{f_{_{0}}(z), f_{_{1}}(z),\cdots,f_{_{m}}(z) \}$ as defined above. Then for every $a(z)$ $\in$ Tor$_{_{i_{j}}}(C)$,  $deg\big(a(z)\big) \geq t_{_{j}}$. Also, Tor$_{_{i_{j}}}(C)=\langle \overline{h_{_{j}}(z)}\rangle$ and $t_{_{j}}$ is the $i_{_{j}}^{th}$ torsional degree of $C$.
        \end{lemma}

        \begin{remark}[\cite{17}]\label{torsion codes containment}
         Let $C=\langle f_{_{0}}(z), f_{_{1}}(z),\cdots,f_{_{m}}(z)\rangle$ be a cyclic code having length $n$ over $R,$ where $f_{_{j}}(z)$ for $j=0,1,\cdots,m$ are polynomials as defined above. Clearly,
          \begin{itemize}
            \item[$(i)$] Tor$_{_{0}}(C)=$Tor$_{_{1}}(C)= \cdots =$Tor$_{_{i_{_{m}}-1}}(C)=\{0\},$
        	\item[$(ii)$] Tor$_{_{i_{_{j}}}}(C) =$ Tor$_{_{i_{_{j}}+1}}(C)= \cdots   =$Tor$_{_{i_{_{j-1}}-1}}(C) \subset $Tor$_{_{i_{_{j-1}}}}(C) \ for \ j=1, 2,  \cdots, m,$
        	\item[$(iii)$] Tor$_{_{i_{_{0}}}}(C) =$ Tor$_{_{i_{_{0}}+1}}(C)= \cdots =$ Tor$_{_{\nu-2}}(C)=$Tor$_{_{\nu-1}}(C).$
          \end{itemize}
        \end{remark}

        \begin{remark}
         For a cyclic code $C$ with generating set as defined above, the above remark  implies that 
         \begin{itemize}
           \item[$(i)$] For $i_{_{0}} \leq i \leq \nu-1,$ the $i^{th}$ torsional degree of $C$ is  $t_{_{0}},$
           \item[$(ii)$] For $1 \leq j \leq m$ and $i_{_{j}} \leq i \leq i_{_{j-1}}-1,$ the $i^{th}$ torsional degree of $C$ is $t_{_{j}}.$
         \end{itemize}
        \end{remark}

        \begin{theorem}[\cite{17}]\label{cardinality}
        	Let $C$ be a cyclic code having arbitrary length $n$ over $R$ generated by polynomials $f_{_{0}}(z), f_{_{1}}(z),\cdots,f_{_{m}}(z)$ as defined earlier. If $\lvert \top \rvert=p^{s}$, then
        	$$\lvert C \rvert = p^{s\big(n\nu-(ni_{_{m}} + t_{_{0}}k_{_{0}}+t_{_{1}}k_{_{1}}+\cdots+t_{_{m}}k_{_{m}})\big)},$$
        	where $t_{_{j}}$ for $j=0, 1, \cdots, m$ are the torsional degrees of Tor$_{_{i_{j}}}(C)$,  $k_{_{0}}=\nu-i_{_{0}}$ and $k_{_{j}}=i_{_{j-1}}-i_{_{j}}$ for $j=1,2,\cdots,m$.
        \end{theorem}

        \begin{theorem}[\cite{17}]\label{hamming distance}
        	Let $C=\langle f_{_{0}}(z), f_{_{1}}(z),\cdots,f_{_{m}}(z)\rangle$ be a cyclic code as defined above. Then $d_{_{H}}(C)=d_{_{H}}\big($Tor$_{_{i_{0}}}(C)\big)=d_{_{H}}\big(\langle\overline{h_{_{0}}(z)}\rangle\big).$
        \end{theorem}

          \begin{remark}\label{division algorithm}
          	Let $\gamma^{i}k(z), \gamma^{j}w(z) \in R[z]$ such that $i \geq j,$ $deg\big(k(z)\big)\geq deg\big(w(z)\big)$ and $w(z)$ is monic. Let $a,b \in R$ be the leading  coefficient of $k(z)$ and $w(z)$ respectively. Then
          	\begin{equation}\label{remainder}
          	\gamma^{i}k(z)-\gamma^{i-j}ab^{-1}z^{deg\big(k(z)\big)-deg\big(w(z)\big)}\gamma^{j}w(z)=\gamma^{i}s_{_{1}}(z),
          	\end{equation}
          	for some $s_{_{1}}(z) \in R[z]$ such that $deg\big(s_{_{1}}(z)\big) < deg\big(k(z)\big).$ If $deg\big( s_{_{1}}(z) \big) \geq deg\big( w(z) \big),$ then applying similar arguement as above on $\gamma^{i}s_{_{1}}(z)$, we have 
          	$$\gamma^{i}s_{_{1}}(z)-\gamma^{i-j}lc\big(s_{_{1}}(z)\big)b^{-1}z^{deg\big( s_{_{1}}(z) \big) - deg\big( w(z) \big)} \gamma^{j}w(z) = \gamma^{i}s_{_{2}}(z),$$
          	for some $s_{_{2}}(z) \in R[z]$ such that $deg\big( s_{_{2}}(z) \big) < deg\big( s_{_{1}}(z) \big).$ Again if $deg\big(s_{_{2}}(z)\big) \geq deg\big(w(z)\big),$ then repeatedly applying the above arguement a finite number of times to obtain polynomials $s_{_{3}}(z), s_{_{4}}(z), \cdots, s_{_{l}}(z)$ in $R[z]$ with $deg\big(s_{_{2}}(z)\big) > deg\big(s_{_{3}}(z)\big) > \cdots > deg\big(s_{_{l}}(z)\big) \geq deg\big(w(z)\big)$ such that 
          	$$\gamma^{i}s_{_{l}}(z)-\gamma^{i-j}lc\big(s_{_{l}}(z)\big)b^{-1}z^{deg\big(s_{_{l}}(z)\big)-deg\big(w(z)\big)}\gamma^{j}w(z)=\gamma^{i}s(z),$$
          	where $s(z) \in R[z]$ and $deg\big(s(z)\big) < deg\big(w(z)\big).$ Now, back substituting all these values of $\gamma^{i}s_{_{l}}(z), \gamma^{i}s_{_{l-1}}(z),\cdots,\gamma^{i}s_{_{1}}(z)$ one by one to finally get that
          	$$\gamma^{i}k(z)-q(z)\gamma^{j}w(z)=\gamma^{i}s(z)$$
          	for $q(z),s(z) \in R[z],$ $deg\big(q(z)\big) \leq deg\big(k(z)\big)-deg\big(w(z)\big)$ and $deg\big(s(z)\big) < deg\big(w(z)\big).$
          \end{remark}

           In the following theorem, a unique set of generators for a cyclic code $C$ over a finite chain ring $R$ has been obtained, which retains all the properties as that of the generating set obtained in \cite{16}.\\
           For a positive integer $t,$ define
           $\mathfrak{B}_{_{t}}=\{ a(z) \in \top[z] \text{ such that } deg\big( a(z) \big) < t \}.$

        \begin{theorem}\label{unique form of genrators}
        	Let $C=\langle f_{_{0}}(z), f_{_{1}}(z), \cdots, f_{_{m}}(z) \rangle$ be a cyclic code having arbitrary length over $R$ as defined above. Then, there exist polynomials $\mathfrak{u}_{_{0}}(z), \mathfrak{u}_{_{1}}(z), \cdots, \mathfrak{u}_{_{m}}(z)$ in $C$ such that for $0 \leq j \leq m,$ 
        	$$\mathfrak{u}_{_{j}}(z)=\sum_{l=i_{_{j}}}^{\nu-1}\gamma^{l}b_{_{j,l}}(z),$$
        	where $b_{_{j,l}}(z) \in \top[z]$ for $i_{_{j}} \leq l \leq \nu-1,$
        	$b_{_{j,i_{_{j}}}}(z)=\overline{h_{_{j}}(z)}$ such that $\overline{h_{_{j}}(z)}$ is the generator polynomial of $i_{_{j}}^{th}$ torsion code of $C$ and  $deg\big(b_{_{j,i_{_{j}}}}(z)\big)=t_{_{j}}.$
        	Further, $b_{_{j,l}}(z) \in \mathfrak{B}_{_{t_{_{j}}}}$ for $i_{_{j}} < l < i_{_{j-1}},$ $b_{_{j,l}}(z) \in \mathfrak{B}_{_{t_{_{r}}}}$ for $i_{_{r}} \leq l < i_{_{r-1}}$ and $j-1 \geq r \geq 1$ and $b_{_{j,l}}(z) \in \mathfrak{B}_{_{t_{_{0}}}}$ for $i_{_{0}} \leq l \leq \nu-1.$
        	Also, $C$ is generated by the set $\{ \mathfrak{u}_{_{0}}(z), \mathfrak{u}_{_{1}}(z), \cdots, \mathfrak{u}_{_{m}}(z) \}$  which retains all the properties as that of the generating set $\{f_{_{0}}(z),f_{_{1}}(z),\cdots,f_{_{m}}(z)\}$ and $\mathfrak{u}_{_{i}}(z)$ are unique in this form.
        \end{theorem}

        \begin{proof}
        	Let $C=\langle f_{_{0}}(z), f_{_{1}}(z), \cdots, f_{_{m}}(z) \rangle$ be a cyclic code over $R$ such that $f_{_{j}}(z)$ are polynomials as defined above. By construction, it is clear that $f_{_{0}}(z)$ is unique in $C.$ Therefore, $f_{_{0}}(z)=\mathfrak{u}_{_{0}}(z).$ Now consider, 
        	\begin{equation}
        		f_{_{1}}(z)=\sum_{l=i_{_{1}}}^{\nu-1}a_{_{1,l}}(z),
        	\end{equation}
        	where $a_{_{1,l}}(z) \in \top[z]$ for $i_{_{1}} \leq l \leq \nu-1,$ $a_{_{1,i_{_{1}}}}(z)=\overline{h_{_{1}}(z)}$ such that $\overline{h_{_{1}}(z)}$ is the generator polynomial of $i_{_{1}}^{th}$ torsion code of $C,$ $deg\big( a_{_{1,i_{_{1}}}}(z) \big) =t_{_{1}}$ and $a_{_{1,l}}(z) \in \mathfrak{B}_{_{t_{_{1}}}}$ for $i_{_{1}} < l \leq \nu-1.$ If $a_{_{1,l}}(z) \in \mathfrak{B}_{_{t_{_{0}}}}$ for $i_{_{0}} \leq l \leq \nu-1,$ then $f_{_{1}}(z)$ is of the desired form. Otherwise, suppose $k \leq \nu-i_{_{0}}-1$ be least non-negative integer such that $a_{_{1,i_{_{0}}+k}}(z) \notin \mathfrak{B}_{_{t_{_{0}}}}.$ Then, $deg\big( a_{_{1,i_{_{0}}+k}} \big) \geq t_{_{0}}.$ By Remark $\ref{division algorithm},$ 
        	\begin{equation}
        	 \gamma^{i_{_{0}}+k}a_{_{1,i_{_{0}}+k}}(z)=\gamma^{i_{_{0}}}h_{_{0}}(z) q^{(1)}_{_{k}}(z) + \gamma^{i_{_{0}}+k}s^{(1)}_{_{k}}(z),
        	\end{equation}
        	for some polynomials $q^{(1)}_{_{k}}(z), s^{(1)}_{_{k}}(z) \in R[z]$ such that $deg\big( s^{(1)}_{_{k}}(z) \big) < t_{_{0}}.$ Let $\gamma^{i_{_{0}}+k}s^{(1)}_{_{k}}(z)= \sum_{l=i_{_{0}}+k}^{\nu-1} \gamma^{l}s^{(1)}_{_{k,l}}(z),$ where $s^{(1)}_{_{k,l}}(z) \in \mathfrak{B}_{_{t_{_{0}}}}$ for $i_{_{0}}+k \leq l \leq \nu-1.$ Substitue this in Equation $(3)$ and then back substitue the value of $\gamma^{i_{_{0}}+k}a_{_{1,i_{_{0}}+k}}(z)$ in Equation $(2)$ to get 
        	$$f_{_{1}}(z)=\sum_{l=i_{_{1}}, ~ l \neq i_{_{0}}+k}^{\nu-1} \gamma^{l}a_{_{1,l}}(z) + \gamma^{i_{_{0}}}h_{_{0}}(z)q^{(1)}_{_{k}}(z) + \sum_{l=i_{_{0}}+k}^{\nu-1}\gamma^{l} s^{(1)}_{_{k,l}}(z).$$
        	This implies 
        	$$f_{_{1}}(z)- \gamma^{i_{_{0}}}h_{_{0}}(z)q^{(1)}_{_{k}}(z) = \sum_{l=i_{_{1}}}^{i_{_{0}}+k-1} \gamma^{l}a_{_{1,l}}(z) + \gamma^{i_{_{0}}+k}s^{(1)}_{_{k,i_{_{0}}+k}}(z) + \sum_{l=i_{_{0}}+k+1}^{\nu-1} \gamma^{l} \big( a_{_{1,l}}(z)+s^{(1)}_{_{k,l}}(z) \big).$$
        	Clearly, the term with content $\gamma^{i_{_{0}}+k}$ on right hand side of the above equation now belongs to $\mathfrak{B}_{_{t_{_{0}}}}.$ Following the same arguments as above for every $a_{_{1,i_{_{0}}+k'}}(z) \notin \mathfrak{B}_{_{t_{_{0}}}}$ where $k < k' \leq \nu-i_{_{0}}-1,$ we can obtain a polynomial say $\mathfrak{u}_{_{1}}(z)=\sum_{l=i_{_{1}}}^{\nu-1} b_{_{1,l}}(z)$ in $C$ by substracting a suitable multiple of $f_{_{0}}(z)$ from $f_{_{1}}(z)$ which will satisfy all the desired properties, i.e., $b_{_{1,l}}
        	(z) \in \top[z],$ $b_{_{1,i_{_{1}}}}(z)=\overline{h_{_{1}}(z)},$ $b_{_{1,l}}(z) \in \mathfrak{B}_{_{t_{_{1}}}}$ for $i_{_{1}} < l < i_{_{0}}$ and $b_{_{1,l}}(z) \in \mathfrak{B}_{_{t_{_{0}}}}$ for $i_{_{0}} \leq l \leq \nu-1$ such that $C=\langle \mathfrak{u}_{_{0}}(z), \mathfrak{u}_{_{1}}(z), f_{_{2}}(z), \cdots, f_{_{m}}(z) \rangle.$ \\
        	Now, consider the polynomial
        	\begin{equation}
        	f_{_{2}}(z)=\sum_{i_{_{2}}}^{\nu-1} a_{_{2,l}}(z),
        	\end{equation}
        	where $a_{_{2,l}}(z) \in \top[z]$ for $i_{_{1}} \leq l \leq \nu-1,$ $a_{_{2,i_{_{2}}}}(z)=\overline{h_{_{2}}(z)},$ $deg\big(a_{_{2,i_{_{2}}}}(z)\big) = t_{_{2}},$ $a_{_{2,l}} \in \mathfrak{B}_{_{t_{_{2}}}}$ for $i_{_{2}} < l \leq \nu-1.$ Further, if $a_{_{{2,l}}}(z) \in \mathfrak{B}_{_{t_{_{1}}}}$ for $i_{_{1}} \leq l < i_{_{0}}$ and $a_{_{2,l}}(z) \in \mathfrak{B}_{_{t_{_{0}}}}$ for $i_{_{0}} \leq l \leq \nu-1,$ then $f_{_{2}}(z)$ is of the desired form. Otherwise, let there exist least positive integers $k \leq \nu-i_{_{0}}-1$ and $r \leq i_{_{0}}-i_{_{1}}-1$ such that $a_{_{2,i_{_{0}}+k}}(z) \notin \mathfrak{B}_{_{t_{_{0}}}}$ and $a_{_{2,i_{_{1}}+r}}(z) \notin \mathfrak{B}_{_{t_{_{1}}}}.$ Using Remark $\ref{division algorithm}$ for $a_{_{2,i_{_{0}}+k}}(z)$ and $a_{_{2,i_{_{1}}+r}}(z),$ we have that 
        	$$\gamma^{i_{_{0}}+k}a_{_{2,i_{_{0}}+k}}(z)-\gamma^{i_{_{0}}}h_{_{0}}(z)q^{(2)}_{_{k}}(z) = \gamma^{i_{_{0}}+k}s^{(2)}_{_{k}}(z)$$
        	$$\text{ and  }\gamma^{i_{_{1}}+r}a_{_{2,i_{_{1}}+r}}(z)-\gamma^{i_{_{1}}}h_{_{1}}(z)q^{(2)}_{_{r}}(z) = \gamma^{i_{_{1}}+r}s^{(2)}_{_{r}}(z)$$
        	such that $q^{(2)}_{_{k}}(z),s^{(2)}_{_{k}}(z),q^{(2)}_{_{r}}(z),s^{(2)}_{_{r}}(z) \in R[z]$ and degree of $s^{(2)}_{_{k}}(z)$ and degree of $s^{(2)}_{_{r}}(z)$ are strictly less than $t_{_{0}}$ and $t_{_{1}},$ respectively. Let $s^{(2)}_{_{k}}(z)=\sum_{l=i_{_{0}}+k}^{\nu-1}\gamma^{l}s^{(2)}_{_{k,l}}(z)$ and $s^{(2)}_{_{r}}(z)=\sum_{l=i_{_{1}}+r}^{\nu-1}\gamma^{l}s^{(2)}_{_{r,l}}(z)$ for  every $s^{(2)}_{_{k,l}}(z),s^{(2)}_{_{r,l}}(z)$ in $\top[z].$ Then, $s^{(2)}_{_{k,l}}(z) \in \mathfrak{B}_{_{t_{_{0}}}}$ for $i_{_{0}} \leq l \leq \nu-1$ and $s^{(2)}_{_{r,l}}(z) \in \mathfrak{B}_{_{t_{_{1}}}}$ for $i_{_{1}}+r \leq l \leq \nu-1.$ Using this to obtain the value of $a_{_{2,i_{_{0}}+k}}(z)$ and $a_{_{2,i_{_{1}}+r}}(z)$ and then back substituting these values in the summand for $f_{_{2}}(z),$ we get
        	$f_{_{2}}(z) -\gamma^{i_{_{1}}}h_{_{1}}(z)q^{(2)}_{_{r}}(z) - \gamma^{i_{_{0}}}h_{_{0}}(z)q^{(2)}_{_{k}}(z) =\sum_{l=i_{_{2}}}^{i_{_{1}}+r-1}\gamma^{l}a_{_{2,l}}(z) + \gamma^{i_{_{1}}+r}s^{(2)}_{_{r,i_{_{1}}+r}}(z) + \sum_{l=i_{_{1}}+r+1}^{i_{_{0}}+k-1}\gamma^{l}\big( a_{_{2,l}}(z) + s^{(2)}_{_{r,l}}(z) \big) + \gamma^{i_{_{0}}+k}s^{(2)}_{_{k,i_{_{0}}+k}}(z) + \sum_{l=i_{_{0}}+k+1}^{\nu-1} \gamma^{l}\big( a_{_{2,l}}(z) + s^{(2)}_{_{r,l}}(z) + s^{(2)}_{_{k,l}}(z) \big).$ Clearly, on the right hand side of this equation, the term with content $\gamma^{i_{_{1}}+r}$ now has degree strictly less than $t_{_{1}}$ and the term with content $\gamma^{i_{_{0}}+k}$ has degree strictly less than $t_{_{0}}.$ Following the similar arguments as above for every $k < k' \leq \nu-i_{_{0}}-1$ and $r < r' \leq i_{_{0}}-i_{_{1}}-1,$ we can finally obtain a polynomial $\mathfrak{u}_{_{2}}(z)=\sum_{l=i_{_{2}}}^{\nu-1}\gamma^{l}b_{_{2,l}}(z)$ in $C$ by substracting a suitable multiple of $f_{_{0}}(z)$ and $f_{_{1}}(z)$ from $f_{_{2}}(z)$ and $\mathfrak{u}_{_{2}}(z)$ satisfies all the desired properties.
        	Similarly, for every $3 \leq j \leq m,$ we can obtain a polynomial $\mathfrak{u}_{_{j}}(z)=\sum_{l=i_{_{j}}}^{\nu-1} b_{_{j,l}}(z)$ in $C$ by substracting suitable multiples of $f_{_{j-1}}(z), f_{_{j-2}}(z),\cdots,f_{_{0}}(z)$ from $f_{_{j}}(z),$ such that $\mathfrak{u}_{_{j}}(z)$ is of the desired form and $C=\langle \mathfrak{u}_{_{0}}(z),\mathfrak{u}_{_{1}},\cdots,\mathfrak{u}_{_{m}}(z) \rangle.$
        	It is clear from the above arguements that these $\mathfrak{u}_{_{j}}(z)$ have same structural properties as that of $f_{_{j}}(z),$ for every $0 \leq j \leq m.$

            Next, we show that the polynomials $\mathfrak{u}_{_{j}}(z),$ $0 \leq j \leq m$ obtained above are unique in this form. Let if possible, $C=\langle \mathfrak{u}_{_{0}}(z),\mathfrak{u}_{_{1}}(z),\cdots,\mathfrak{u}_{_{m}}(z) \rangle = \langle w_{_{0}}(z),w_{_{1}}(z),\cdots,w_{_{m}}(z) \rangle,$ where $\mathfrak{u}_{_{j}}(z)= \sum_{l=i_{_{j}}}^{\nu-1} \gamma^{l} b_{_{j,l}}(z)$ and $w_{_{j}}(z)=\sum_{l=i_{_{j}}}^{\nu-1} \gamma^{l} d_{_{j,l}}(z),$ such that $b_{_{j,l}}(z), d_{_{j,l}}(z) \in \top[z]$ for $i_{_{j}} \leq l \leq \nu-1,$
            $b_{_{j,i_{_{j}}}}(z)=d_{_{j,i_{_{j}}}}(z)=\overline{h_{_{j}}(z)}$ for the generator polynomial $\overline{h_{_{j}}(z)}$ of the $i_{_{j}}^{th}$ torsion code of $C$ and  $deg\big(b_{_{j,i_{_{j}}}}(z)\big)=deg\big(d_{_{j,i_{_{j}}}}(z)\big)=t_{_{j}}.$
            Further, $b_{_{j,l}}(z), d_{_{j,l}}(z) \in \mathfrak{B}_{_{t_{_{j}}}}$ for $i_{_{j}} < l < i_{_{j-1}},$ $b_{_{j,l}}(z), d_{_{j,l}}(z) \in \mathfrak{B}_{_{t_{_{r}}}}$ for $i_{_{r}} \leq l < i_{_{r-1}}$ and $j-1 \geq r \geq 1$ and $b_{_{j,l}}(z), d_{_{j,l}}(z) \in \mathfrak{B}_{_{t_{_{0}}}}$ for $i_{_{0}} \leq l \leq \nu-1.$              
            Clearly,
            $w_{_{0}}(z)=f_{_{0}}(z)=\mathfrak{u}_{_{0}}(z)$ by the above construction. For $1 \leq j \leq m,$ consider the polynomial 
            \begin{equation*}
             w_{_{j}}(z)-\mathfrak{u}_{_{j}}(z)= \sum_{l=i_{_{j}}}^{\nu-1} \gamma^{l}\big( d_{_{j,l}}(z)-b_{_{j,l}}(z) \big).
            \end{equation*}
            Let us denote the polynomials $d_{_{j,l}}(z)-b_{_{j,l}}(z)$ by $e_{_{j,l}}(z)$ for $i_{_{j}} \leq l \leq \nu-1.$ Then,
            \begin{equation}\label{E(z)}
            w_{_{j}}(z)-\mathfrak{u}_{_{j}}(z)=\gamma^{i_{_{j}}+1}\sum_{l=i_{_{j}}+1}^{\nu-1} \gamma^{l-i_{_{j}}-1} e_{_{j,l}}(z),
            \end{equation} 
            since $d_{_{j,i_{_{j}}}}(z)=b_{_{j,i_{_{j}}}}(z)=\overline{h_{_{j}}(z)},$ i.e.,   $e_{_{j,i_{_{j}}}}(z)=0.$ We have that $\phi \big( \sum_{l=i_{_{j}}+1}^{\nu-1} \gamma^{l-i_{_{j}}-1} e_{_{j,l}}(z) \big) = e_{_{j,i_{_{j}}+1}}(z) \in Tor_{_{i_{_{j}}+1}}(C).$
            From Remark $\ref{torsion codes containment}$ and Lemma $\ref{torsion code generators},$ we have Tor$_{_{i_{_{j}}+1}}(C)=$Tor$_{_{i_{_{j}}}}(C)=\langle \overline{h_{_{j}}(z)} \rangle$ for $i_{_{j}}+1 < i_{_{j-1}}.$ Therefore, $e_{_{j,i_{_{j}}+1}}(z) \in \langle \overline{h_{_{j}}(z)} \rangle$ but $deg\big( e_{_{j,i_{_{j}}+1}}(z) \big) < t_{_{j}}$ which implies that $e_{_{j,i_{_{j}}+1}}(z)=0.$ Putting this in Equation $(\ref{E(z)})$ and applying the same arguments a finite number of times till we get $e_{_{j,l}}(z)=0$ for $i_{_{j}} \leq l < i_{_{j-1}}.$ Putting this in Equation $(\ref{E(z)}),$ we have 
            $$w_{_{j}}(z)-\mathfrak{u}_{_{j}}(z)= \gamma^{i_{_{j-1}}}\sum_{l=i_{_{j-1}}}^{\nu-1}\gamma^{l-i_{_{j-1}}}e_{_{j,l}}(z).$$ 
            We have 
            $\phi\big( \sum_{l=i_{_{j-1}}}^{\nu-1} \gamma^{l-i_{_{j-1}}} e_{_{j,l}}(z) \big) = e_{_{j,i_{_{j-1}}}}(z) \in Tor_{_{i_{_{j-1}}}}(C).$
            Using Lemma $3.2,$ we get that $e_{_{j,i_{_{j-1}}}}(z) \in \langle \overline{h_{_{j-1}}(z)} \rangle.$ Then $e_{_{j,i_{_{j-1}}}}(z)=0,$ since $deg\big( e_{_{j,i_{_{j-1}}}}(z) \big) < t_{_{j-1}}.$ Using this in Equation $(\ref{E(z)}),$ we get that $$w_{_{j}}(z)-\mathfrak{u}_{_{j}}(z)= \gamma^{i_{_{j-1}}+1}\sum_{l=i_{_{j-1}}+1}^{\nu-1}\gamma^{l-i_{_{j-1}}+1}e_{_{j,l}}(z).$$ Again, we have $\phi\big( \sum_{l=i_{_{j-1}}+1}^{\nu-1} \gamma^{l-i_{_{j-1}}+1} e_{_{j,l}}(z) \big) = e_{_{j,i_{_{j-1}}+1}}(z) \in Tor_{_{i_{_{j-1}}+1}}(C).$ Using Remark $\ref{torsion codes containment}$ and Lemma $\ref{torsion code generators},$ we get that $e_{_{j,i_{_{j-1}}+1}}(z) \in Tor_{_{i_{_{j-1}}}+1}(C)=Tor_{_{i_{_{j-1}}}}(C)= \langle \overline{h_{_{j-1}}(z)} \rangle$ for $i_{_{j-1}}+1 < i_{_{j-2}}.$ Then $e_{_{j,i_{_{j-1}}+1}}(z)=0,$ since $deg\big( e_{_{j,i_{_{j-1}}+1}}(z) \big) < t_{_{j-1}}.$ Putting this in Equation $(\ref{E(z)}),$ and repeatedly applying the same argument a finite number of times till we get $e_{_{j,l}}(z)=0$ for $i_{_{j-1}} \leq l < i_{_{j-2}}.$ Working in a similar manner for every $l \leq \nu-1,$ we can finally conclude that $w_{_{j}}(z)-\mathfrak{u}_{_{j}}(z)=0.$ Hence, the generator polynomials $\mathfrak{u}_{_{j}}(z)$ for $0 \leq j \leq m$ are unique in $C.$	
         \end{proof}

         \begin{remark}
         	It is observed that the unique set of generators obtained in Theorem 3.1 forms a Grobner basis for $C$ over $R.$
         \end{remark}

\section{MDS and MHDR cyclic codes over a finite chain ring}

           In this section, the minimal spanning set and rank of a cyclic code $C$ over a finite chain ring $R$ have been obtained. All MDS (Maximum Distance Separable) cyclic codes of arbitrary length over $R$ have been determined. All MHDR cyclic codes having length which is not coprime to characteristic of residue field of the ring have also been determined.

          \begin{theorem}\label{rank}
          	Let $C$ be a cyclic code having arbitrary length $n$ over a finite chain ring $R.$ Then $rank(C)=n-t_{_{0}},$ where $t_{_{0}}$ is the degree of minimal degree polynomial in $C.$
          \end{theorem}
          
          \begin{proof}
          	Let $C$ be a cyclic code having arbitrary length n over $R.$ Let $\{ \mathfrak{u}_{_{0}}(z), \mathfrak{u}_{_{1}}(z), \cdots, \mathfrak{u}_{_{m}}(z)\},$ be a unique set of generators for $C$ as obtained above. Clearly, the set
          	$S = \{  \mathfrak{u}_{_{m}}(z), z\mathfrak{u}_{_{m}}(z), \cdots, z^{n-t_{_{m}}-1}\mathfrak{u}_{_{m}}(z),
          	 \mathfrak{u}_{_{m-1}}(z), z\mathfrak{u}_{_{m-1}}(z), \cdots,\\ z^{n-t_{_{m-1}}-1}\mathfrak{u}_{_{m-1}}(z),
          	 \cdots, 
          	 \mathfrak{u}_{_{1}}(z), z\mathfrak{u}_{_{1}}(z), \cdots, z^{n-t_{_{1}}-1}\mathfrak{u}_{_{1}}(z),
          	 \mathfrak{u}_{_{0}}(z), z\mathfrak{u}_{_{0}}(z), \cdots,\\ z^{n-t_{_{0}}-1}\mathfrak{u}_{_{0}}(z)\}$
          	spans $C.$ Now, we shall prove that
          	$S' = \{  \mathfrak{u}_{_{m}}(z), z\mathfrak{u}_{_{m}}(z), \cdots, z^{n-t_{_{m}}-1}\mathfrak{u}_{_{m}}(z),
          	 \mathfrak{u}_{_{m-1}}(z), z\mathfrak{u}_{_{m-1}}(z), \cdots, z^{t_{_{m}}-t_{_{m-1}}-1}\mathfrak{u}_{_{m-1}}(z),\\
          	 \cdots, 
          	 \mathfrak{u}_{_{1}}(z), z\mathfrak{u}_{_{1}}(z), \cdots, z^{t_{_{2}}-t_{_{1}}-1}\mathfrak{u}_{_{1}}(z),
          	 \mathfrak{u}_{_{0}}(z), z\mathfrak{u}_{_{0}}(z), \cdots, z^{t_{_{1}}-t_{_{0}}-1}\mathfrak{u}_{_{0}}(z)\}$
          	also spans $C.$ For this, we need to prove that $z^{t_{_{j+1}}-t_{_{j}}}\mathfrak{u}_{_{j}}(z),$ for $0 \leq j \leq m-1$ are in $span$ $S'.$ We shall show this by induction on $j$. First, we prove that $z^{t_{_{1}}-t_{_{0}}}\mathfrak{u}_{_{0}}(z) \in$ $span$ $S'.$ 
          	Clearly, $z^{t_{_{1}}-t_{_{0}}}\mathfrak{u}_{_{0}}(z)$ is a polynomial of degree $t_{_{1}}$ in $C.$ Then, we have $z^{t_{_{1}}-t_{_{0}}}\mathfrak{u}_{_{0}}(z)-\gamma^{i_{_{0}}-i_{_{1}}} \mathfrak{u}_{_{1}}(z)= q_{_{0}}(z) \mathfrak{u}_{_{0}}(z)$ for some $q_{_{0}}(z) \in R[z]$ with degree less than $t_{_{1}}-t_{_{0}}$ which implies that $z^{t_{_{1}}-t_{_{0}}}\mathfrak{u}_{_{0}}(z)-\gamma^{i_{_{0}}-i_{_{1}}} \mathfrak{u}_{_{1}}(z) \in$ $span$ $S'.$ Therefore, we have
          	$z^{t_{_{1}}-t_{_{0}}}\mathfrak{u}_{_{0}}(z) \in$ $span$ $S'.$ Now, suppose that $z^{t_{_{2}}-t_{_{1}}}\mathfrak{u}_{_{1}}(z), z^{t_{_{3}}-t_{_{2}}}\mathfrak{u}_{_{2}}(z), \cdots, z^{t_{_{j}}-t_{_{j-1}}}\mathfrak{u}_{_{j-1}}(z) \in$ $span$ $S'$ for $1 \leq j \leq m-1.$  Now, we will show that $z^{t_{_{j+1}}-t_{_{j}}}\mathfrak{u}_{_{j}}(z) \in$ $span$ $S'.$ Clearly, $z^{t_{_{j+1}}-t_{_{j}}}\mathfrak{u}_{_{j}}(z)$ is a polynomial of degree $t_{_{j+1}}$ in $C.$ Then, $z^{t_{_{j+1}}-t_{_{j}}}\mathfrak{u}_{_{j}}(z)-\gamma^{i_{_{j}}-i_{_{j+1}}}\mathfrak{u}_{_{j+1}}(z) \in \langle \mathfrak{u}_{_{0}}(z), \mathfrak{u}_{_{1}}(z),\cdots, \mathfrak{u}_{_{j}}(z)\rangle.$ Then, $z^{t_{_{j+1}}-t_{_{j}}}\mathfrak{u}_{_{j}}(z)=\gamma^{i_{_{j}}-i_{_{j+1}}}\mathfrak{u}_{_{j+1}}(z)+m_{_{0}}(z)\mathfrak{u}_{_{0}}(z)+m_{_{1}}\mathfrak{u}_{_{1}}(z) +\cdots+m_{_{j}}\mathfrak{u}_{_{j}}(z),$ where $m_{_{i}}(z) \in R[z]$ and $deg\big(m_{_{i}}(z)\big) < t_{_{i+1}}-t_{_{i}}$ for all $i$, $0 \leq i \leq j.$ This implies that $m_{_{i}}\mathfrak{u}_{_{i}}(z) \in$ $ span $ $S'$ for $0 \leq i \leq j$ which further implies that $z^{t_{_{j+1}}-t_{_{j}}}\mathfrak{u}_{_{j}}(z) \in$ $span$ $S'.$ Therefore, we have $z^{t_{_{j+1}}-t_{_{j}}}\mathfrak{u}_{_{j}}(z) \in$ $span$ $S'$ for all $j,$ $0 \leq j \leq m-1.$

          	Next, we prove linear independence of $S'.$ Let if possible, there exist $\alpha_{_{j,r}} \in R$ such that 
          	\begin{align} \label{eqn6}
          	 z^{n-t_{_{m}}-1} \mathfrak{u}_{_{m}}(z) &= \alpha_{_{m,0}}\mathfrak{u}_{_{m}}(z) + \alpha_{_{m,1}} z \mathfrak{u}_{_{m}}(z) + \cdots + \alpha_{_{m,n-t_{_{m}}-2}} z^{n-t_{_{m}}-2} \mathfrak{u}_{_{m}}(z) \nonumber \\
          	& + \alpha_{_{m-1,0}} \mathfrak{u}_{_{m-1}}(z) + \alpha_{_{m-1,1}} z \mathfrak{u}_{_{m-1}}(z)+ \cdots \nonumber \\
          	& + \alpha_{_{m-1,t_{_{m}}-t_{_{m-1}}-1}} z^{t_{_{m}}-t_{_{m-1}}-1} \mathfrak{u}_{_{m-1}}(z) + \cdots \nonumber \\
          	& + \alpha_{_{1,0}} \mathfrak{u}_{_{1}}(z) + \alpha_{_{1,1}} z \mathfrak{u}_{_{1}}(z) + \cdots + \alpha_{_{1,t_{_{2}}-t_{_{1}}-1}} z^{t_{_{2}}-t_{_{1}}-1} \mathfrak{u}_{_{1}}(z) \nonumber \\
          	& + \alpha_{_{0,0}} \mathfrak{u}_{_{0}}(z) + \alpha_{_{0,1}} z \mathfrak{u}_{_{0}}(z) + \cdots + \alpha_{_{0,t_{_{1}}-t_{_{0}}-1}} z^{t_{_{1}}-t_{_{0}}-1} \mathfrak{u}_{_{0}}(z).
          	\end{align}
          	This implies that $z^{n-t_{_{m}}-1} \mathfrak{u}_{_{m}}(z) = \alpha_{_{m}}(z)\mathfrak{u}_{_{m}}(z)+\alpha_{_{m-1}}(z)\mathfrak{u}_{_{m-1}}(z)+\cdots+\alpha_{_{0}}(z)\mathfrak{u}_{_{0}}(z),$ where $\alpha_{_{m}}(z)=\alpha_{_{m,0}}+\alpha_{_{m,1}}z+\cdots+\alpha_{_{m,n-t_{_{m}}-2}}z^{n-t_{_{m}}-2}$ and $\alpha_{_{i}}(z)=\alpha_{_{i,0}}+\alpha_{_{i,1}}z+\cdots+\alpha_{_{i,t_{_{i+1}}-t_{_{i}}-1}}z^{t_{_{i+1}}-t_{_{i}}-1}$ for $0 \leq i \leq m-1.$ Clearly, $deg\big( \alpha_{_{m}}(z) \big) \leq n-2$ and $deg\big( \alpha_{_{i}}(z) \big) \leq t_{_{i+1}}-1$ for all $i,$ $0 \leq i \leq m-1.$ Multiplying Equation $(\ref{eqn6})$ by $\gamma^{\nu-i_{_{m-1}}},$ we get
          	\begin{equation}
          	\label{eqn7}
          	z^{\nu-t_{_{m}}-1}\gamma^{\nu-i_{_{m-1}}}\mathfrak{u}_{_{m}}(z) = \alpha_{_{m}}(z)\gamma^{\nu-i_{_{m-1}}}\mathfrak{u}_{_{m}}(z).
          	\end{equation}
         	Then degree of $LHS$ of Equation $(\ref{eqn7})$ is $n-1$ but that of $RHS$ is atmost $n-2$ which is a contradiction. Therefore, $z^{n-t_{_{m}}-1} \mathfrak{u}_{_{m}}(z)$ can not be expressed as a linear combination of elements of $S'.$  We can apply similar arguments to prove that  none of $z^{t_{_{m}}-t_{_{m-1}}-1} \mathfrak{u}_{_{m-1}}(z), z^{t_{_{m-1}}-t_{_{m-2}}-1} \mathfrak{u}_{_{m-2}}(z), \cdots, z^{t_{_{1}}-t_{_{0}}-1} \mathfrak{u}_{_{0}}(z)$ can be expressed as a linear combination of elements of $S'.$ Therefore, we get that $S'$ is linearly independent and hence it is a $minimal$ $spanning$ set for $C.$ It follows that           	
          	$rank(C)  = n - t_{_{0}}.$
          \end{proof}

           The following theorem determines all the MDS cyclic codes of arbitrary length over a finite chain ring $R$.

           \begin{theorem}\label{mds}
           	A cyclic code $C$ having length $n$ over $R$ is MDS if and only if it is principally generated by a monic polynomial and Tor$_{_{0}}(C)$ is an MDS cyclic code having length $n$ over $\top$ with respect to Hamming metric. 
           \end{theorem}

           \begin{proof}
           	Let $C=\langle \mathfrak{u}_{_{0}}(z), \mathfrak{u}_{_{1}}(z), \cdots, \mathfrak{u}_{_{m}}(z) \rangle$ be an MDS cyclic code having length $n$ over $R$ such that $\mathfrak{u}_{_{j}}(z),$ $0 \leq j \leq m$ are polynomials as in Theorem $\ref{unique form of genrators}$. Because $C$ is MDS, therefore $\lvert C \lvert = \lvert R \lvert ^{n-d_{_{H}}(C)+1}.$ Using Theorem $\ref{cardinality},$ we have $p^{s\big(n\nu-ni_{_{m}} - t_{_{0}}k_{_{0}}-t_{_{1}}k_{_{1}}-\cdots-t_{_{m}}k_{_{m}}\big)}=p^{s\nu\big(n-d_{_{H}}(C)+1\big)}$ which  implies that $ni_{_{m}} +t_{_{0}}k_{_{0}}+t_{_{1}}k_{_{1}}+\cdots+t_{_{m}}k_{_{m}} = \nu \big( d_{_{H}} (C)-1 \big).$ Form here, we can conclude that $t_{_{j}}=0$ for $1 \leq j \leq m$ and $i_{_{m}}=0$ because $i_{_{m}}+k_{_{0}}+k_{_{1}}+\cdots+k_{_{m}} = \nu$ and $t_{_{m}} > t_{_{m-1}} > \cdots > t_{_{0}} \geq d_{_{H}}(C)-1.$ This implies that $C$ is principally generated by a monic polynomial and $t_{_{0}}=d_{_{H}}(C)-1.$ Using Theorem$ \ref{cardinality}$ and Theorem $\ref{hamming distance},$ we have $\lvert \top \rvert^{\big( n-d_{_{H}}(Tor_{_{0}}(C))+1 \big) }=p^{s\big( n-d_{_{H}}(Tor_{_{0}}(C))-1 \big)}=p^{s\big( n-d_{_{H}}(C)-1 \big)}=p^{s(n-t_{_{0}})}=\lvert Tor_{_{0}}(C) \rvert.$ Thus,$Tor_{_{0}}(C)$ is an MDS cyclic code over the residue field $\top. $
           	
           	Conversely, suppose a cyclic code $C$ having length $n$ over $R$ is principally generated by a monic polynomial, say $\mathfrak{u}_{_{0}}(z)$ as obtained in Theorem $\ref{unique form of genrators}$ and $Tor_{_{0}}(C)$ is an MDS code over $\top.$ This means that $i_{_{0}}=0$ and $\lvert Tor_{_{0}}(C) \rvert = \lvert \top \rvert^{\big( n-d_{_{H}}(Tor_{_{0}}(C))+1 \big)}.$ Using Theorem $\ref{cardinality}$ and Theorem $\ref{hamming distance},$ we can conclude that $ \lvert R \lvert ^{n-d_{_{H}}(C)+1}=p^{s \nu \big( n-d_{_{H}}(Tor_{_{0}}(C))+1 \big)}=p^{s\nu(n-t_{_{0}})}=\lvert C \lvert,$ i.e., $C$ is an MDS cyclic code over $R.$
           \end{proof}

        The following lemma by A. Sharma and T. Sidana determines the $hamming$ $distance$ of a cyclic code $C$ of length $n'p^{r},$ $(n',p)=1$ and $r \geq 1$ over a finite chain ring $R$ as given in \cite{23}. 
        
        \begin{lemma}[\cite{23}]\label{anuradha dhc}
        	Let $C$ be a cyclic code having length $n=n'p^{r},$ for $(n',p)=1$ and $r \geq 1$ over $R$. Then 
        	\begin{equation}
        	d_{_{H}}(C)=
        	\begin{cases}
        	1, & \text{if}\ \  t_{_{0}}=0 \\
        	l+2, & \text{if}\ \ lp^{r-1} +1 \leq t_{_{0}} \leq (l+1)p^{r-1}\\ 
        	\ \ \ \ & \text{with}\ \ 0 \leq l \leq p-2\\
        	(i+1)p^{k}, & \text{if}\ \ p^{r}-p^{r-k}+(i-1)p^{r-k-1}+1 \leq t_{_{0}} \leq p^{r}-p^{r-k}+ip^{r-k-1}\\
        	\ \ \ \ & \text{with}\ \ 1 \leq i \leq p-1 \text{and}\ \ 1 \leq k \leq r-1\\
        	\end{cases}
        	\end{equation}
        \end{lemma}

       We use Lemma $\ref{anuradha dhc}$ above to determine all MHDR cyclic codes of length $n'p^{r},$ $(n',p)=1$ and $r \geq 1$ over $R$ in Theorem $\ref{mhdr p}$ and $\ref{mhdr nps}$ below.
        
        \begin{theorem}\label{mhdr p}
        	A cyclic code $C$ of length  $n'p,$ $(n',p)=1$ over a finite chain ring $R$ is an MHDR code.
        \end{theorem}
        
        \begin{proof}
        	Let $C$ be a cyclic code of length $n'p,$ $(n',p)=1$ over $R.$ By Lemma $\ref{anuradha dhc}$, we have
        	\begin{equation*}
        	d_{_{H}}(C)=
        	\begin{cases}
        	1, & \text{if}\ \  t_{_{0}}=0 \\
        	t_{_{0}}+1, & \text{if}\  1 \leq t_{_{0}}\leq p-1\\ 
        	\end{cases}
        	\end{equation*}
        	which implies that $d_{_{H}}(C)=t_{_{0}}+1=n-rank(C)+1$ for $0 \leq t_{_{0}} \leq p-1$ using Theorem $\ref{rank}.$ Hence, a cyclic code of length $n'p,$ $(n',p)=1$ over $R$ is always an MHDR code.
        \end{proof}

        \begin{theorem}\label{mhdr nps}
        	Let $C$ is a cyclic code having length $n=n'p^{r}, r > 1$ over $R.$ Then $C$ is MHDR if and only if $t_{_{0}} \in \{0,1,p^{r}-1\}.$
        \end{theorem}
    
        \begin{proof}
        	By Lemma $\ref{anuradha dhc},$ we have
        	\begin{itemize}
        		\item[$(i)$] for $t_{_{0}}=0,$ the Hamming distance of $C$ is $1$ which is same as $n-rank(C)+1$ using Theorem $\ref{rank}.$ So, $C$ is an MHDR code.
        		
        	    \item[$(ii)$] for $lp^{r-1} +1 \leq t_{_{0}} \leq (l+1)p^{r-1}$ with $0 \leq l \leq p-2,$ the Hamming distance of $C$ is $l+2.$ Here, $C$ is MHDR if and only if $d_{_{H}}(C)=n-rank(C)+1,$ i.e., $l+1=t_{_{0}}$ using Theorem $\ref{rank}.$ Then, $lp^{r-1} +1 \leq t_{_{0}}$ would imply $lp^{r-1} +1 \leq l+1,$ i.e., $l(p^{r-1} -1) \leq 0.$ It follows that $l(p^{r-1} -1) = 0$  which implies $l=0,$ since $p^{r-1} \neq 1.$ Then, $C$ is MHDR if and only if $t_{_{0}}=1.$
        	    
        	    \item[$(iii)$] for $k=r-1,$ $t_{_{0}}=p^{r}-p+i, 1 \leq i \leq p-1,$ the Hamming distance of $C$ is $(i+1)p^{r-1}.$ $C$ is an MHDR code if  and only if $(i+1)p^{r-1}=n-rank(C)+1=t_{_{0}}+1$ using Theorem $\ref{rank}.$ Then, we have  
        	    $p^{r}-p+i=t_{_{0}}=(i+1)p^{r-1}-1.$ It follows that $p(p^{r-1}-1)=(i+1)(p^{r-1}-1)$
        	    which implies $i=p-1,$ since $p^{r-1}\neq 1.$ Then, $C$ is MHDR for $t_{_{0}}=p^{r}-1.$ It can be easily seen that for other values of $t_{_{0}},$ $C$ is not an MHDR code.
        	    \end{itemize}
        \end{proof}

          \begin{theorem}
          	Let $C$ be an MDS cyclic code having arbitrary length over $R.$ Then $C$ is also an MHDR code over $R.$
          \end{theorem}
      
          \begin{proof}
          	Let $C$ be an MDS cyclic code having arbitrary length $n$ over $R.$ By Theorem $\ref{mds}$, $C$ is principally generated by a monic polynomial over $R$ say $\mathfrak{u}_{_{0}}(z)$ with degree $t_{_{0}}$ and $i_{_{0}}=0$ and Tor$_{_{0}}(C)$ is also an MDS code over $\top.$ Then 
          	\begin{equation}\label{9}
          	 \lvert Tor_{_{0}}(C) \rvert = p^{s(n-d_{_{H}}(C)+1)}.
          	\end{equation}
          	Also, from Theorem $\ref{cardinality},$ we have
          	\begin{equation}\label{10}
          	\lvert Tor_{_{0}}(C) \rvert = p^{s(n-t_{_{0}})}.
          	\end{equation}
          	Equation $(\ref{9})$ and $(\ref{10})$ together with Theorem $\ref{rank}$ imply that $d_{_{H}}(C)=t_{_{0}}+1=n-rank(C)+1.$
          	Therefore, $C$ is an MHDR cyclic code over $R.$ 
          \end{proof}

          However, Example $\ref{example of mhdr not mds}$ shows that converse of the above statement is not true.
          
          \begin{example}\label{example of mhdr not mds}
          	Let $R=Z_{_{5}}+5Z_{_{5}}$. Let $C=\langle 5, (z-1)^{24} \rangle$ be a cyclic code having length $n=25$ over $R.$ Here, $i_{_{0}}=1, i_{_{1}}=0, t_{_{0}}=0,$ $t_{_{1}}=24,$ $rank(C)=25$ and $d_{_{H}}(C)=1.$ $C$ is an MHDR cyclic code over $R$ (using Theorem $\ref{mhdr nps}$). However, $C$ is not an MDS code, since it is not principally generated (using Theorem $\ref{mds}$). 
          \end{example}

          \begin{example}
          	Let $R=Z_{_{5}}+5Z_{_{5}}.$ Let $C=\langle (z-1)^{24} \rangle$ be a cyclic code having length $n=25$ over $R.$ Here, $i_{_{0}}=0,$ $t_{_{0}}=24,$ $rank(C)=1$ and $d_{_{H}}(C)=24.$ $C$ is an MHDR cyclic code over $R$ (using Theorem $\ref{mhdr nps}$). Also, $C$ is an MDS code, since it is principally generated by a monic polynomial and $\lvert Tor_{_{0}}(C) \rvert=5=\lvert Z_{_{5}}\rvert^{n-d_{_{H}}(Tor_{_{0}}(C))+1}$ (using Theorem $\ref{mds}$). 
          \end{example}

          \begin{example}
          	Let $R=Z_{_{2}}+\gamma Z_{_{2}}+\gamma^{2}Z_{_{2}}+\gamma^{3}Z_{_{2}}.$ Let $C=\langle (z^{2}-1)+\gamma(z-1) + \gamma^{2}(z-1) + \gamma^{3} \rangle$ be a cyclic code having length $n=6$ over $R.$ Here, $i_{_{0}}=0,$ $t_{_{0}}=2,$ $rank(C)=4$ and $d_{_{H}}(C)=3.$ It is principally generated by a monic polynomial and $\lvert Tor_{_{0}}(C) \rvert=2^{4}=\lvert Z_{_{2}}\rvert^{n-d_{_{H}}(Tor_{_{0}}(C))+1}=2^{6-3+1}=2^{4},$ so $C$ is an MDS code over $R$ by using Theorem $\ref{mds}.$ Also, from Theorem $\ref{mhdr p},$ $C$ is also an MHDR code.
          \end{example}

          \begin{example}
          	Let $R=Z_{_{2}}+\gamma Z_{_{2}}+\gamma^{2}Z_{_{2}}+\gamma^{3}Z_{_{2}}.$ Let $C=\langle \gamma^{2}(z^{3}-1) + \gamma^{3}(z^{2}-1) \rangle$ be a cyclic code having length $n=6$ over $R.$ Here, $i_{_{0}}=2,$ $t_{_{0}}=3,$ $rank(C)=3$ and $d_{_{H}}(C)=2.$ It is not generated by a monic polynomial, so by Theorem $\ref{mds}$, $C$ is not an MDS code. Also, from Theorem $\ref{mhdr p}$, $C$ is not an MHDR code.
          \end{example}

          \begin{example}
          	Let $R=Z_{_{3}}+\gamma Z_{_{3}}+\gamma^{2}Z_{_{3}}.$ Let $C=\langle \gamma^{2}(z^{2}-1), \gamma(z^{2}-1)^{3} + \gamma^{2}(z-1) \rangle$ be a cyclic code having length $n=18$ over $R.$ Here, $i_{_{0}}=2,$ $i_{_{1}}=1$ $t_{_{0}}=2,$ $t_{_{1}}=6,$ $rank(C)=16$ and $d_{_{H}}(C)=2.$ It is not generated by a monic polynomial, so by Theorem $\ref{mds}$, it is not an MDS code. Also, from Theorem $\ref{mhdr nps}$, $C$ is not an MHDR code. 
          \end{example}

\section{Conclusion}

            In this work, a unique set of generators for a cyclic code having arbitrary length over a finite chain ring with arbitrary nilpotency index has been established. The minimal spanning set and rank of the code have also been determined. Further, sufficient as well as necessary conditions for a cyclic code having arbitrary length to be an MDS code and for a cyclic code having length which is not coprime to characteristic of residue field of the ring, to be an MHDR code have been obtained. Some examples of optimal cyclic codes have also been presented.

\section*{Acknowledgements}

         The first author gratefully acknowledges the support provided by the Council of Scientific and Industrial Research (CSIR), India in the form of a research fellowship.

\end{document}